\documentclass[conference]{IEEEtran}

\usepackage{booktabs} 

\usepackage{amsmath}
\usepackage{amsfonts}
\usepackage{multirow}

\usepackage{multicol}
\usepackage{graphicx,dblfloatfix}
\usepackage[ruled, linesnumbered]{algorithm2e}
\usepackage{xcolor}
\usepackage[numbers,square,comma,compress,sort]{natbib}
\usepackage{booktabs}

\begin{document}

\title{\vspace{-8mm} Machine Learning-based Low Overhead Congestion Control Algorithm for Industrial NoCs \vspace{-6mm}
}
\author{Shruti Yadav Narayana$^1$, Sumit K. Mandal$^{2}$,  Raid Ayoub$^3$, Michael Kishinevsky$^3$, Umit Y. Ogras$^1$ \\
$^1$Dept. of ECE, University of Wisconsin-Madison; 
$^2$Dept. of CSA, Indian Institute of Science, Bangalore, India; \\ 
$^3$Intel Corporation, Hillsboro, OR \vspace{-9mm}
}

\maketitle

\begin{abstract}



Network-on-Chip (NoC) congestion builds up during heavy traffic load and cripples the system performance by stalling the cores.
Moreover, congestion leads to wasted link bandwidth due to blocked buffers and bouncing packets.
Existing approaches throttle the cores after congestion is detected, reducing efficiency and wasting line bandwidth unnecessarily.
In contrast, we propose a lightweight machine learning-based technique that helps predict congestion in the network.
Specifically, our proposed technique collects the features related to traffic at each destination.
Then, it labels the features using a novel time reversal approach.
The labeled data is used to design a low overhead and an explainable decision tree model used at runtime congestion control.
Experimental evaluations with synthetic and real traffic on industrial 6$\times$6 NoC show that the proposed approach increases fairness and memory read bandwidth by up to 114\% with respect to existing congestion control technique while incurring less than 0.01\% of overhead.


\end{abstract}


\vspace{-2mm}
\section{Introduction}
\vspace{-1mm}
	


Systems-on-chip (SoCs) with multi-core processors use networks-on-chip (NoCs) for fast and energy-efficient communication between the processing elements.
NoCs consist of three main components: 
1) routers, 2) links, and 3) queues for storing the packets. 
Buffered NoCs, such as those with wormhole routing, store the flits that make up the packets in intermediate routers~\cite{marculescu2008outstanding}. 
In contrast, bufferless NoCs, commonly used in industrial processors, store the packets only at the endpoints, i.e., in the egress queues of the traffic sources and ingress queues of the destinations. 
Under heavy traffic, i.e., when the rate of incoming packets is high, finite size queues fill up and apply backpressure.
NoCs implement backpressure mechanisms to prevent packet losses.
For example, when the ingress queues at the destination are full, 
the buffered NoCs stall the packet and propagate the backpressure upstream. 
In contrast, bufferless industrial NoCs deflect the packets to one of the available output ports. Hence, the deflected packets use NoC resources until they return to the deflection point~\cite{doweck2017inside}.
In either case, the backpressure eventually propagates all the way back to the traffic sources (i.e., the cores) and prevents them from injecting new packets into the NoC.
As a result, the throughput (the number of packets processed per unit time) decreases, and the overall performance of the SoC deteriorates.

To address the congestion problem, researchers have proposed congestion control mechanisms for industrial NoCs~\cite{tam2018skylake}.
This technique monitors the queue occupancy at the sinks.
If the queue occupancy exceeds a predetermined threshold, 
a \textit{distress on} signal is sent to the sources.
The source then stops sending packets until 
it receives a \textit{distress off} signal, 
which is triggered after queue occupancy drops below a certain threshold.
\begin{figure}[t]
	\centering
	\includegraphics[width=0.85\columnwidth]{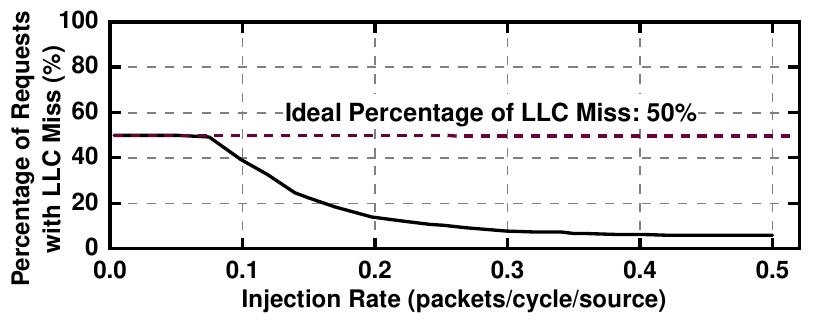}
	\vspace{-3mm}
	\caption{Percentage of miss packets with state-of-the-art congestion control technique. The traffic is generated with 50\% LLC miss rate (shown in red dashed line). However, the percentage of miss packets decreases to 7\% at the highest injection rate which is extremely unfair to miss traffic.}
	\vspace{-6mm}
	\label{fig:src_throttle_target}
\end{figure}
This mechanism reduces the NoC congestion and limits the packet latency in the NoC.
However, the resulting throughput is lower under a heavier workload.
Moreover, the mechanism is not fair towards the requests that experience a miss in the shared last level cache (LLC).
For example, Figure~\ref{fig:src_throttle_target} shows the percentage of traffic with LLC miss as a function of the traffic injection rate.
The percentage of requests with LLC miss denotes the proportion of the requests fetched from the memory controller.
In this experiment, we generate synthetic traffic that will result in 50\% (the dotted line on the figure) LLC miss rate.
At low traffic loads, the observed percentage of LLC miss rate is 50\%, as expected.
However, the percentage of completed transactions with LLC misses drops as the traffic intensity increases and becomes as low as 7\% when the network becomes heavily congested.
Similar unfairness happens also at other LLC miss rates.
In addition, static techniques depend on the predetermined threshold that does not change with the workload, while reactive techniques act after congestion happens~\cite{akbar2021novel, zhao2020fine}.
Hence, they do not maintain the throughput of the NoC under heavy traffic.


The goal of this work is twofold. First, it aims to maximize and sustain the memory read/write bandwidth provided to the processing cores. Achieving this goal ensures that the cores use the NoC bandwidth effectively and that their performance does not degrade with NoC congestion. Second, it aims to maximize the fairness between the LLC hit and miss traffic. Without this goal, the requests with an LLC miss experience starvation (as demonstrated in Figure 1) since they have much longer delays than those with LLC hit. 
We propose a proactive congestion control (a.k.a., source throttling) technique to achieve these goals. The first step of the proposed approach is a supervised learning framework enabled by a novel design of experiments and time reversal techniques. The second step designs a lightweight decision tree using the data from these experiments. This decision tree determines whether any given sink node will likely experience congestion or not (before the queue is blocked). Finally, the decision tree is used at runtime to control the traffic sources, i.e., the cores. If a sink is likely congested, the cores stop sending new requests to that sink until the congestion signal is cleared. Experimental evaluations with synthetic and realistic traces show that the proposed technique increases memory read bandwidth by up to 114\% and the percentage of missed traffic by up to 3.1$\times$ compared to a state-of-the-art congestion control technique.


\textit{The major contributions of the work are as follows:}
\begin{itemize}
    \item A novel time reversal approach and supervised learning to construct a decision tree for NoC congestion control,
    \item End-to-end congestion control algorithm for industrial NoCs,
    \item Thorough experimental evaluations showing up to 114\% higher memory read bandwidth than a state-of-the-art technique with less than 0.01\% of overhead.
\end{itemize}

The rest of this paper is organized as follows.
Prior work related to congestion control in NoC are discussed in Section~\ref{sec:rel_work}.
We review the background of the work in Section~\ref{sec:background}.
The proposed technique is described in detail in Section~\ref{sec:method}.
The experimental results are presented in Section~\ref{sec:expt_eval}.
Finally, Section~\ref{sec:concl} concludes the paper.

\vspace{-2mm}
\section{Related Work} \label{sec:rel_work}



Existing NoC congestion control techniques can be broadly classified as 1) Global and 2) Local.
The global congestion control techniques assess the congestion status of the whole network.
Depending on the congestion status, the packet injection rates of all the sources are regulated~\cite{jain1998congestion, smai1998global,marculescu2008outstanding}.
In contrast, local congestion control techniques monitor the congestion at each node~\cite{tam2018skylake, jeffers2016intel, zhao2020fine, akbar2021novel}.
State-of-the-art industrial NoCs monitor the ingress queue sizes of each node~\cite{tam2018skylake, jeffers2016intel}.
If the size exceeds a certain threshold, then the injection of packets from all the sources is stopped.
The packet injection resumes when the occupancies of all the queues drop below another predetermined threshold. With this technique, congestion at any of the ingress queues leads to throttling at all sources, leading to conservative behavior.
Authors in~\cite{zhao2020fine} propose a fine-grained source throttling method for NoCs with mesh topology.
In this work, the routers which are most affected by congestion are identified.
Then, these routers are used to estimate the NoC congestion status.
A heterogeneous congestion criterion for 2D mesh-NoC is proposed in~\cite{akbar2021novel}.
When NoC congestion occurs in a node, the packets whose trajectory is through the congested node are stalled in the source.
However, all the congestion control techniques described above are reactive i.e., the congestion criteria kicks in only after the congestion physically occurs in the NoC.


Proactive congestion control techniques for NoCs are proposed in~\cite{ogras2006prediction, wang2019ann}.
The technique proposed in~\cite{ogras2006prediction} estimates the availability of the neighboring router through analytical expression.
When a traffic source observes that the input port connected to it does not have availability then it does not send the packets.
Authors in~\cite{wang2019ann} propose an artificial neural network (ANN)-based global admission controller for NoC.
In this work, the admission controller slows down the injection rate from the sources by factor determined by the ANN.
The aforementioned techniques are applicable to an NoC where the packets can wait at each router on its path.
However, industrial NoCs are priority aware and incorporates deflection routing where the packets already injected in the NoC can never stop.
Therefore, existing proactive congestion controls techniques are not applicable to industrial NoCs.
In contrast, we propose a proactive congestion control technique to increase memory read/write bandwidth and to improve the fairness between LLC hit and miss traffic for industrial NoCs with deflection routing.
To the best of our knowledge, this is the first proactive congestion control technique proposed for industrial NoCs with deflection routing.

\vspace{-3mm}
\section{System Architecture and Background} \label{sec:background}

\begin{figure}[t]
	\centering
	\vspace{-2mm}
 	\includegraphics[width=0.7\linewidth]{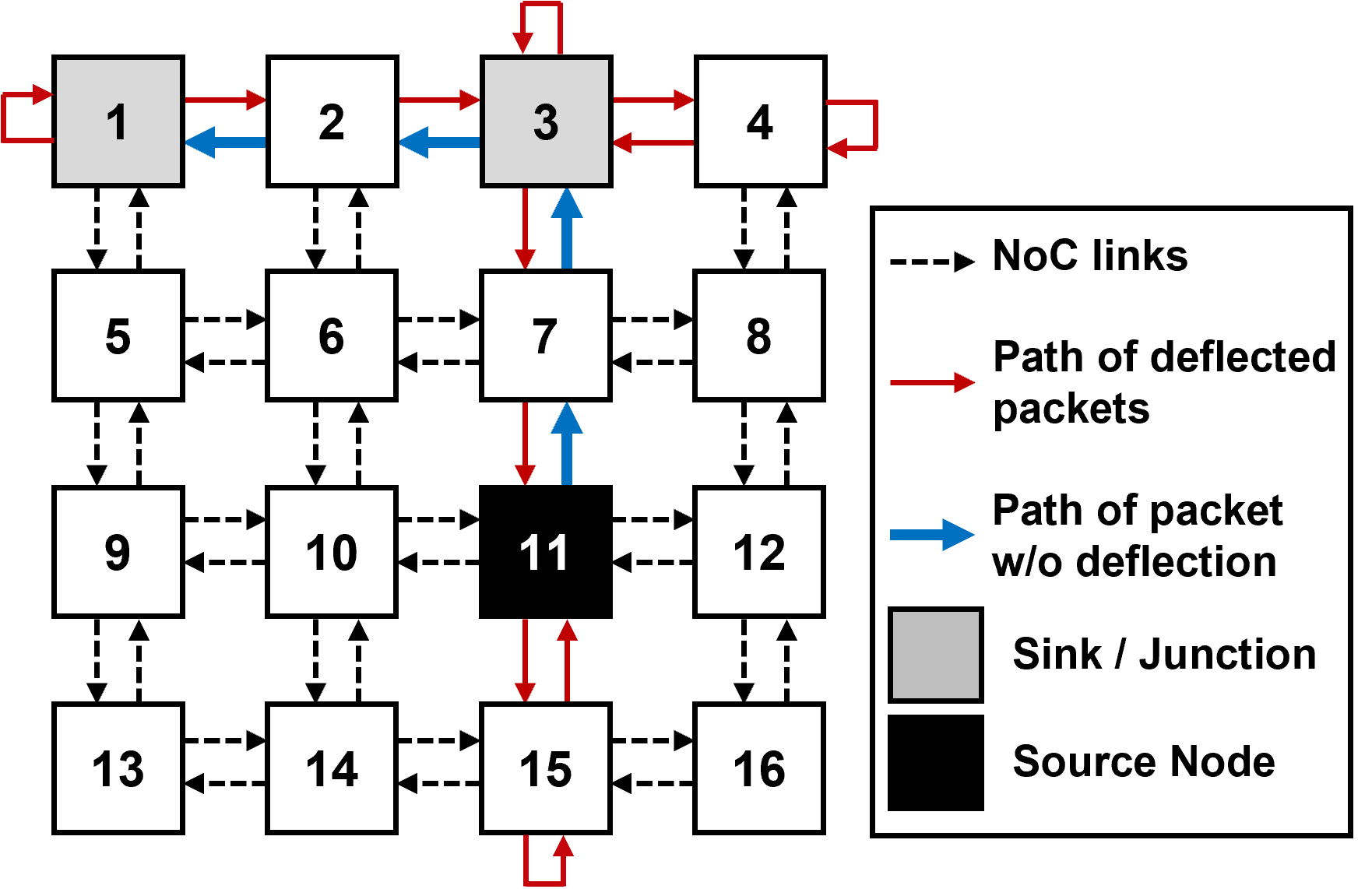}
  \vspace{-3mm}
	\caption{A representative 4$\times$4 mesh-NoC with deflection routing.}
   \vspace{-6mm}
	\label{fig:bnc_expl}
\end{figure}

 \vspace{-2mm}
\subsection{NoCs with Deflection Routing}
\vspace{-2mm}

This work targets NoCs used in high-end servers and state-of-the-art many core architectures~\cite{jeffers2016intel}.
Figure~\ref{fig:bnc_expl} shows a 4$\times$4 mesh NoC architecture, each column of which is also used in client systems, such as Intel i7 processors~\cite{rotem2015intel}. 
Hence, the proposed congestion control technique is applicable to a wide range of priority-aware industrial NoCs, where the packets already in the network have higher priority than the packets waiting in the egress queues of the sources.
Assume that Node 11 in Figure~\ref{fig:bnc_expl} sends a packet to Node 1 following Y-X routing (highlighted by thick blue arrows).
Deflection in priority-aware NoCs happens when the queue at the turning point (Node 3) or final destination (Node 1) becomes  full. This can happen if the receiving node, such as a cache controller, cannot process the packets fast enough. The probability of observing a full queue increases with smaller queues (needed to save area) and heavy traffic load from the cores. 
If the packet is deflected at the destination node, it circulates within the same row (the red thin arrows), as shown in Figure~\ref{fig:bnc_expl}.
Consequently, a combination of regular and deflected traffic can load the corresponding row and pressure the queue at the turning point (Node 3). This, in turn, can lead to deflection on the column which propagates the congestion towards the source wasting useful NoC bandwidth. 

\vspace{-3mm}
\subsection{Background on Cache Coherency Flow}
\vspace{-1mm}

This work assumes a local L1/L2 cache at each node, a distributed LLC, and non-inclusive MESI-like cache-coherency flow~\cite{papamarcos1984low}.
If a request from a core is not present in L1 or L2 cache, the request is sent to LLC.
If the request is present in the LLC, then the corresponding data is returned from LLC to the requesting core.
If the request is not present in the LLC then the request is forwarded to the memory controller.
The corresponding data is fetched from the memory controller and returned to the core.
The proposed congestion control technique is independent of the number of cores, LLC banks and on-chip memory controllers.


\section{ML-Based Proactive Source Throttling} \label{sec:method}
\vspace{-1mm}

\subsection{Overview of the Approach}
\vspace{-2mm}



When packets in the NoC are deflected at the sink, they continue to use the NoC bandwidth and aggravate congestion.
Hence, the proposed runtime technique works as follows:

\begin{enumerate}
    \item It monitors the congestion indicators, i.e., the features of our machine learning (ML) model, at each sink queue and determines whether they are likely to be blocked,
    \item If a given queue is likely to be blocked, it sets a congestion signal at that sink. Otherwise, it clears the congestion signal.
    \item The sources check the congestion signal at the destination before sending a new request. If the congestion signal is set, they throttle the corresponding request and move on to the next request. 
    The requests to the sinks with congestion signal are delayed until the congestion signal is cleared. 
\end{enumerate}

We note that checking the congestion signal at the destination does not incur any additional overhead compared to existing techniques~\cite{tam2018skylake, doweck2017inside}, since they also incorporate similar mechanism. They also have a small ($\sim$ 10 cycles in our case) deterministic delay when the distress information is carried by a simple dedicated time-division-multiplexed channel.

The fundamental question is to determine when to throttle a source. Since the network traffic dynamics are fast, time-dependent, and nonlinear, we need to consider not only the current queue occupancies but also first and second order factors that can lead to congestion. For example, consider an ingress queue with depth 32. An occupancy of 16 packets may be safe if the current input traffic rate is lower than the service rate and the level of burstiness is low. Since the average occupancy is likely to be decreasing as time progresses, the sources do not need to be throttled. In contrast, an occupancy of 16 packets may be dangerous if the average occupancy is increasing. 
Therefore, a holistic approach must consider all relevant features summarized in Section~\ref{sec:features}. 
Moreover, determining the optimal criteria is non-trivial even when all the features are available. 
Hence, the second component of the proposed approach is to design a decision tree using an innovative data collection and labelling technique presented in Section~\ref{sec:DT-design}. 
Finally, the last step is implementing the lightweight controller that uses the congestion signals and local criteria to throttle the source (Section~\ref{sec:throttling}).

\vspace{-3mm}
\subsection{Features used for Supervised Learning} \label{sec:features}
\vspace{-1mm}


To construct the machine learning-based model, we first collect the dataset required for training.
The dataset consists of features ($\mathcal{F}$) listed in Table~\ref{tab:features} with corresponding labels ($\mathcal{L}$).
Here $\mathcal{F} = (f^1, f^2, ..., f^N)$, where $N$ is the number of features, $f^j \in \mathbb{R}$, $1 \leq j \leq N$ and $\mathcal{L} \in \{0, 1\}$.
The features ($\mathcal{F}$) are sampled every time a packet arrives at the ingress queue at the sink. 
If the queue is not full, the packet is written to the queue. Otherwise, the packet bounces. 
Sampling the features in both conditions (sink or bounce) enables us to monitor congestion accurately at sink node.

To capture the features accurately, we compute exponentially weighted moving average (EWMA) of each feature as: 
\begin{equation} \label{eq:ewma}
    \bar f_i^j = \alpha f_i^j + (1-\alpha) \bar f_{i-1}^j, ~ i > 0, ~ 1 \leq j \leq N
\end{equation}
In this equation, $\bar f_i^j$ denotes EWMA of the feature $f^j$ for $i^\mathrm{th}$ packet, $f_i^j$ denotes the original value of the feature $f^j$ (e.g. injection rate) and $\alpha$ is the degree of mixing parameter ($0 \leq \alpha \leq 1$).
The value of $\alpha$ is tuned to track the average accurate without a significant delay.
The feature values are smoothened over time by computing EWMA.
We implemented EWMA computation in a cycle-accurate industrial simulator.
A five point derivative is computed for the features involving gradient.
We track all the features in Table~\ref{tab:features}, which are potentially useful for congestion control.
Since data collection is an offline process, the EWMA computation overheads are inconsequential.
After deploying the machine learning model, EWMA of only the selected features are tracked at runtime.

\begin{table}[t]
\setlength\tabcolsep{0.75pt}
\caption{List of features collected at each sink.}
\vspace{-2mm}
\footnotesize
\begin{tabular}{|l|l|}
\hline
Injection rate to the sink queue                                                              & Total injection rate (sunk + deflected)  \\ \hline
\begin{tabular}[c]{@{}l@{}}Co-eff. of variation of\\ the total traffic (sunk + deflected)\end{tabular}  & \begin{tabular}[c]{@{}l@{}}Co-eff. of variation of inter-arrival\\ time of the traffic to the sink queue\end{tabular} \\ \hline
Rate of deflected packets                                                       & Mean service time of the sink queue                \\ \hline
\begin{tabular}[c]{@{}l@{}}Co-eff. of variation of\\ deflected packet inter-arrival time \end{tabular}                                                      & \begin{tabular}[c]{@{}l@{}}Co-eff. of variation of\\ sink queue inter-departure time \end{tabular}                \\ \hline
Occupancy                                                                                     & Probability that the sink queue is full                                                                                    \\ \hline
Gradient of injection rate                                                                    & Gradient of queue occupancy                                                                                                      \\ \hline
\begin{tabular}[c]{@{}l@{}}Gradient of total\\(sunk + deflected) injection rate\end{tabular} &   Gradient of probability of sink being full                                                                                                                       \\ \hline
\end{tabular}
\label{tab:features}
\end{table}

\vspace{-1mm}
\subsection{Training Data Collection and Decision Tree} \label{sec:DT-design}
\vspace{-1mm}


\noindent\textbf{Labeling the features:}
The collected features indicate the ingress queue and NoC congestion state at sampling time. For example, the features will capture if a sink queue is full and deflects a packet. 
However, one must throttle the source before the queue becomes full, i.e., before the onset of congestion. 
The main challenge is to know that a packet will bounce before it is even injected into the network. 
Having this knowledge at runtime is impossible, but we mitigate this challenge using a \textbf{\textit{novel time reversal approach}} described next.
The generation time stamps of all the deflected packets are recorded while sampling the features. If a packet is  deflected at the sink,  we know that the source must have been throttled at the generation time of this packet.
This sense of time in our comprehensive simulation data enables us to go back to the generation time of the deflected packet and label the collected features around that time accordingly.

\begin{figure}[t]
	\centering
	\includegraphics[width=1\columnwidth]{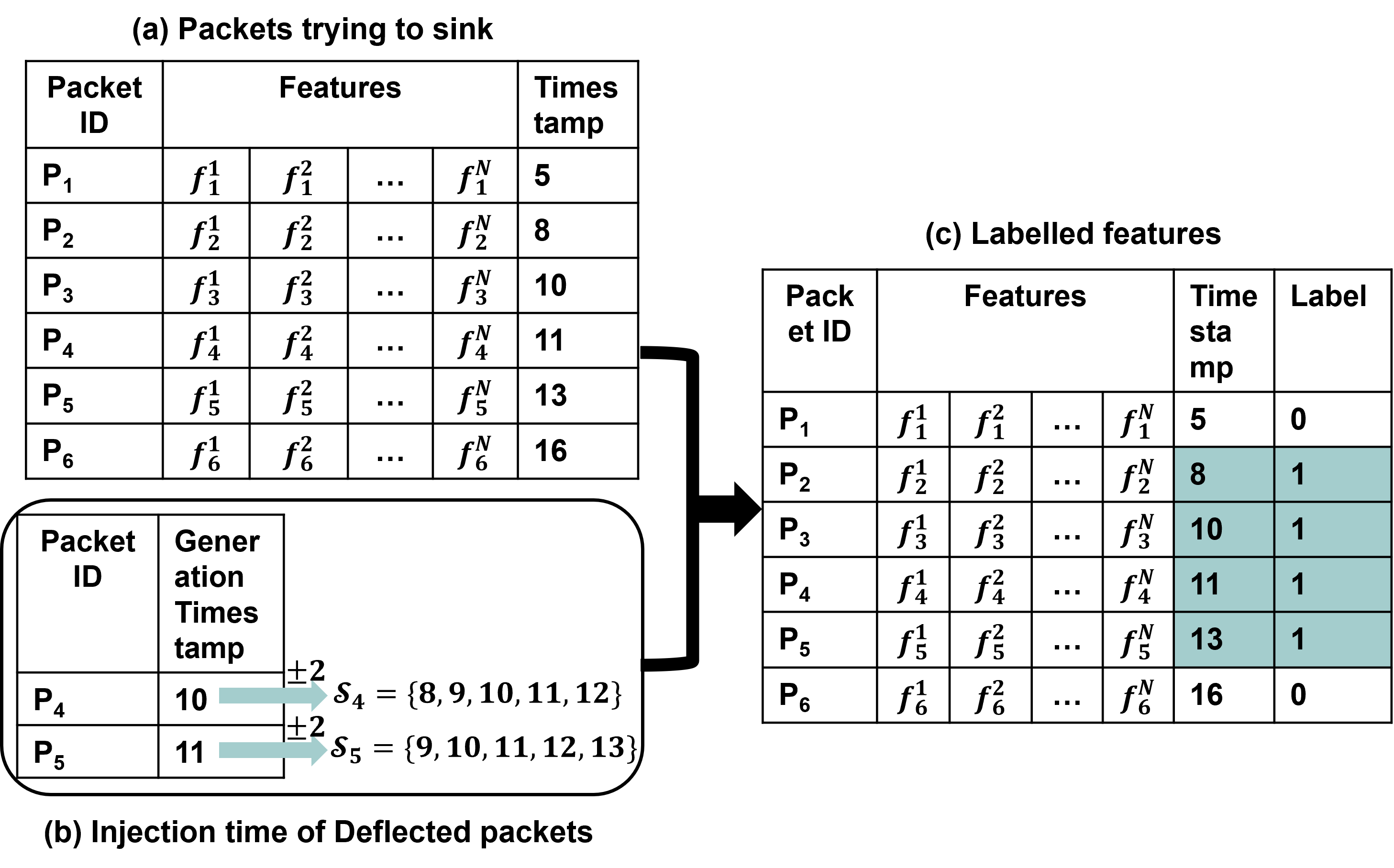}
    \vspace{-8mm}
	\caption{An illustrative example of our proposed time reversal approach to label the features.}
    \vspace{-6mm}
	\label{fig:labeling}
\end{figure}

Figure~\ref{fig:labeling} shows an illustrative example of our proposed time reversal approach for labelling the features.
Figure~\ref{fig:labeling}(a) shows the features sampled for six packets arriving at the ingress queue.
Along with the features, the timestamps when the packets attempted to sink are also sampled (last column of the table).
Apart from sampling the features of the packets arriving at the ingress queue, we also sample the generation timestamps of the deflected packets.
As shown in the Figure~\ref{fig:labeling}(b), there are two deflected packets -- $P_4$ and $P_5$.
The generation timestamps ($d_j$ in Equation~\ref{eq:label}) when they were injected from the source are 10 and 11 respectively.
Next, we compute a set of timestamps for each deflected packets (two in this case) which are within 2 cycles of the injection timestamps.
From $P_4$, we get $\mathcal{S}_4 = \{ 8,9,10,11,12 \}$ and from $P_5$, we get $\mathcal{S}_5 = \{ 9,10,11,12,13 \}$.
If $t_i$ is the timestamp of the packet $P_i$, where $1 \leq i \leq 5$, then we label $P_i$ as 1 if $t_i \in \mathcal{S}_4 \cup \mathcal{S}_5$.
If $d_j$ is the generation timestamp of when the $j^\mathrm{th}$ deflected packet and $t_i$ is the timestamp of the $i^\mathrm{th}$ packet arriving at the sink, in general we label ($l_i$) the features of the $i^\mathrm{th}$ packet arriving at the sink as:
\begin{equation} \label{eq:label}
l_i = 
\begin{cases}
1, & \text{if $(d_j - \Delta) \leq t_i \leq (d_j + \Delta)$ } \\
0,  & \text{Otherwise}
\end{cases}
\end{equation}
where $\Delta = 2$ in this example.
A label of 0 denotes that if the source sends packet to that particular sink, then it \textit{will not result in congestion}.
A label of 1 denotes that if the source sends packet to that particular sink, then it \textit{will result in congestion}.
Therefore, all the features with timestamp within the range of $\Delta$ ($\Delta > 0$) of $d_i$ are labelled as 1.
In other words, features within a range of $\Delta$ timestamps from the same timestamp as the generation timestamp of the deflected packets are labelled as 1.
The features of the packets with label of 1 are highlighted in Figure~\ref{fig:labeling}(c).

\noindent\textbf{Supervised Learning:} We can employ any supervised learning algorithm to create a model which can take congestion control decision.
In this work, we choose binary decision tree since decision tree incurs low hardware overhead (detailed in Section~\ref{sec:hw_ov}).
The output of the decision tree is either 0 or 1.
An output of 0 denotes that cores can send packet without congesting the NoC.
An output of 1 denotes that there is a possibility of congestion in the near future and cores should stop injecting packets in the NoC.
We observe that the decision tree obtained through supervised learning supports our idea of proactive congestion control.
For an example, the decision tree returns an output of 1 if both the occupancy of the sink is high and the gradient of injection rate to the sink is positive.

\vspace{-2mm}
\subsection{Local Source Control} \label{sec:throttling}
\vspace{-1mm}

Each source (e.g., the CPU cores) has controller in the NoC interface.
The controller checks the congestion signal from the decision tree at each sink.
Due to deflection routing, the sources which are located at the boundary of floorplan (e.g., Node-1, 5, 9, 13 in Figure~\ref{fig:bnc_expl}) have highest priority.
Therefore, packets sent from these sources do not compete with the packets waiting at other sources with lower priority.
Hence, the sources with highest priority can inject freely and cause congestion.
In addition to the sink nodes, a local condition 
at sources with highest priority can also proactively hint future congestion.
Therefore, we also implement a local condition for the sources with the highest priority. 
Let $N$ be the current occupancy of the destination sink, and 
$N_T$ be the target occupancy.
According to Little's law, $\lambda \times t_{avg}$ more packet can be written to the queue, where $\lambda$ is the injection rate to the ingress and $t_{avg}$ is the average time between two source throttling decisions~\cite{little1961proof}. 
Hence, the traffic source is throttled if $N + \lambda t_{avg} > N_T$, i.e., the queue can become full.


\begin{algorithm}[t]
\caption{End-to-end congestion control algorithm} \label{algo:st_algo}
\small
\SetAlgoLined
\SetNoFillComment
\textbf{Input:} Absolute value of the features, mixing parameter ($\alpha$), size of the sink queue ($N$), target occupancy ($N_T$) \\
\textbf{Output:} To throttle (1) or not to throttle the source (0) \\

$L = LC(N_T, N, \lambda)$ \\
\If {L == 1} {
 return 1 \\
} \Else {
 $\bar {\mathcal{F}} \leftarrow$  EWMA of the features using Equation~\ref{eq:ewma} \\
    $D = DT(\mathcal{\bar {\mathcal{F}}})$ \\
 return $D$ \\
}
\end{algorithm}
\setlength{\textfloatsep}{0pt}
Algorithm~\ref{algo:st_algo} shows the end-to-end algorithm for congestion control which combines the decision from decision tree model and the local condition.
The input to the algorithm is the absolute value of the features, mixing parameter ($\alpha$), occupancy of the sink queue ($N$), and target occupancy ($N_T$).
First, the controller checks the local condition ($LC$).
$LC$ always returns false for the sources with lower priority.
If the local condition's output ($L$) is true, then the algorithm returns true.
Otherwise, EWMA of the features ($\bar {\mathcal{F}} $) are computed following Equation~\ref{eq:ewma}.
Then, the algorithm returns the output of the decision tree ($D$).

\vspace{-3mm}
\section{Experimental Evaluations} \label{sec:expt_eval}



\vspace{-3mm}
\subsection{Experimental Setup}
\vspace{-1mm}

We use a cycle-accurate industrial NoC simulator to evaluate our proposed approach on a 6$\times$6 mesh NoC with two memory controllers. 
The NoC architecture is similar to the one used in recent industrial SoCs~\cite{tam2018skylake, doweck2017inside}.
Due to classified nature of the simulator and architecture, we present normalized values.
Each simulation is run for 600k cycles (with a warm-up period of 100k) to reach steady-state values. The experiments
consider a non-inclusive MESI-like cache-coherency protocol~\cite{papamarcos1984low} with varying traffic and last-level cache (LLC) hit rates. 


\vspace{-3mm}
\subsection{Accuracy of Decision Tree}
\vspace{-1mm}

We perform simulations for LLC hit rate of 0.5 with different injection rates.
The smoothing parameter ($\alpha$ in Equation~\ref{eq:ewma}) is set as $\frac{1}{16}$ and $\Delta$ (in Equation~\ref{eq:label}) is 5.
The entire dataset is divided into 70\% training data and 30\% validation.
Table~\ref{tab:accuracy} shows the accuracy of predicting label-0 and label-1 for the validation data with decision trees having different depths.
As a reminder, label-0 denotes that if the sources inject packets, it will not lead to congestion and vice-versa.
Therefore, if the labeled feature is 0 and the predicted label is 1, the decision tree will unnecessarily stop the cores from injecting packets.
This scenario might be okay since it will not lead to congestion.
However, if the labeled feature is 1 and the predicted label is 0, the core will still inject packets when it should have stopped.
This misprediction will lead to congestion in NoC.
Therefore, the accuracy of predicting label-1 is more important than the accuracy of predicting label-0.
We observe that the decision tree with depth 4 has the highest accuracy in predicting label-1.
The decision tree with a depth lower than 4 has lower prediction accuracy for label-0 and label-1, 
while a deeper decision tree has a lower accuracy for label-1 due to overfitting.
Therefore, we choose the decision tree with depth 4 for evaluation.
We note that, the decision tree for each sink is trained offline (once) and the same decision tree is used for congestion control with any incoming workload.

\begin{table}[t]
\centering
\vspace{-8mm}
\caption{Accuracy(\%) of decision trees with different depths. Decision tree of depth 4 is chosen based on the accuracy.}
\vspace{-2mm}
\begin{tabular}{|c|ccccccc|}
\hline
\multirow{2}{*}{} & \multicolumn{7}{c|}{Decision tree depth}                                                                                                                                     \\ \cline{2-8} 
                  & \multicolumn{1}{l|}{2}    & \multicolumn{1}{l|}{3}    & \multicolumn{1}{l|}{4}    & \multicolumn{1}{l|}{5}    & \multicolumn{1}{l|}{6}    & \multicolumn{1}{l|}{7}    & 8    \\ \hline
Label-0           & \multicolumn{1}{l|}{93.3} & \multicolumn{1}{l|}{93.2} & \multicolumn{1}{l|}{93.7} & \multicolumn{1}{l|}{94.9} & \multicolumn{1}{l|}{96.1} & \multicolumn{1}{l|}{96.2} & 96.5 \\ \hline
Label-1           & \multicolumn{1}{l|}{95.9} & \multicolumn{1}{l|}{97.4} & \multicolumn{1}{l|}{97.6} & \multicolumn{1}{l|}{97.5} & \multicolumn{1}{l|}{96.3} & \multicolumn{1}{l|}{95.6} & 94.8 \\ \hline
\end{tabular}
\vspace{0mm}
\label{tab:accuracy}
\end{table}



 \vspace{-3mm}
\subsection{Comparison of Average Transaction Latency}
\vspace{-1mm}

The primary goal of our congestion control technique is to reduce the number of deflected packets so that there is no wastage of NoC bandwidth.
We observe that when no congestion control is enabled, the rate of deflected packets increases with increasing injection rate.
For example, with an injection rate of 0.27 and LLC hit rate of 0.2, the average rate of deflected packets is 0.08.
In this scenario, our proposed congestion control technique sees no deflected packets.
A reduced number of deflected packets reduces NoC congestion, so packets experience lower wait time and average latency.

\begin{figure}[t]
	\centering
	\vspace{-8mm}
	\includegraphics[width=0.9\columnwidth]{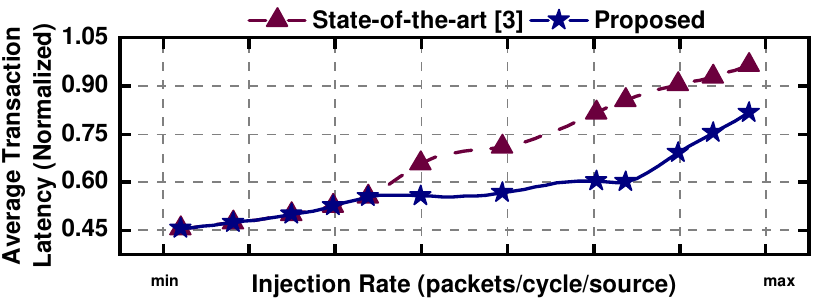}
	\vspace{-3mm}
	\caption{Comparison of average transaction latency for 70\% hit rate. Lower transaction latency indicates less congestion.}
        \vspace{-6mm}
	\label{fig:avg_trans_lat}
\end{figure}

Figure~\ref{fig:avg_trans_lat} shows the comparison of average transaction latency for varying injection rates with LLC hit rate of 70\%.
The comparison is between the congestion control technique used in state-of-the-art industrial NoC~\cite{tam2018skylake} and our proposed approach.
The average transaction latency denotes the round trip latency from the generation of a read/write request to its completion.
In the state-of-the-art congestion control technique, if the occupancy of the sink exceeds a predetermined value, the sources are throttled. The sources restart injecting packets in the NoC if the sink occupancy becomes lower than another predetermined value.
Therefore, the state-of-the-art control technique is reactive.
In contrast, our proposed congestion control technique predicts congestion and provides a proactive decision to throttle the sources.
As a result, compared to the reactive state-of-the-art, our proposed proactive congestion control technique throttles at the onset of congestion in NoC, without wasting memory read bandwidth.
From figure~\ref{fig:avg_trans_lat} it is observed that the proposed congestion control technique reduces the average transaction latency by up to 30\%
compared to the state-of-the-art approach.
We also observe a similar improvement in average transaction latency for other LLC hit rates.
For example, the proposed technique improves the average transaction latency by 7\% for an LLC hit rate of 0.2 on average.

\vspace{-3mm}
\subsection{Comparison of Percentage of LLC Miss}
\vspace{-1mm}

\begin{figure}[t]
	\centering
	\includegraphics[width=0.9\columnwidth]{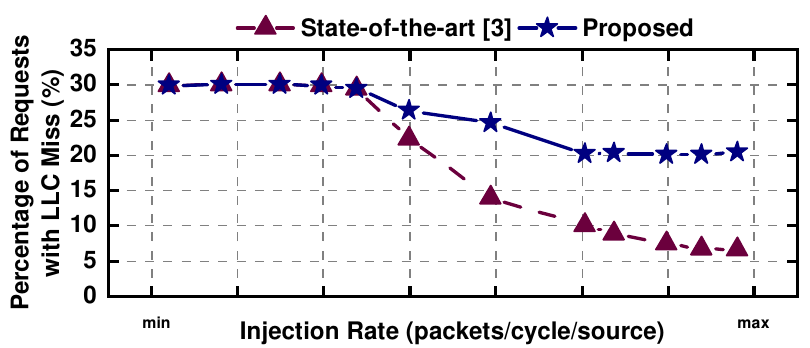}
	\vspace{-3mm}
	\caption{Comparison of percentage of LLC miss for 70\% LLC hit rate (30\% LLC miss). Higher percentage of LLC miss indicates that the congestion control technique is more fair towards the miss traffic.}
	\vspace{0mm}
    \label{fig:miss_perc}
\end{figure}

This section compares the percentage of requests with LLC miss between the state-of-the-art congestion control technique and our proposed technique.
The percentage of requests with LLC miss denotes the proportion of the requests fetched from the memory controller.
Since our proposed technique reduces NoC congestion, more requests with LLC miss are allowed to be fetched from the memory controller.
Therefore, our proposed technique consistently results in a higher percentage of requests with LLC miss, as shown in Figure~\ref{fig:miss_perc} compared to the state-of-the-art congestion control technique.
In this case, the synthetic traffic is generated with a 70\% hit rate, i.e., ideally, 30\% of the traffic should be miss traffic.
We observe that at a lower injection rate, both techniques result in 30\% of requests with LLC miss since there is no congestion in the NoC.
With the increasing injection rate, the percentage of requests with LLC miss reduces.
However, our proposed congestion control technique shows up to 3.1$\times$
improvement in the percentage of requests with LLC miss at the higher injection rate.
We also observe similar LLC miss percentage improvement for other LLC hit rates.
For example, the proposed technique improves the LLC miss percentage by 1.2$\times$ for LLC hit rate of 0.2.
Therefore, the proposed technique is fairer towards the requests with LLC miss than the state-of-the-art technique.

\begin{figure}[t]
	\centering
	 \vspace{-8mm}
	\includegraphics[width=0.9\columnwidth]{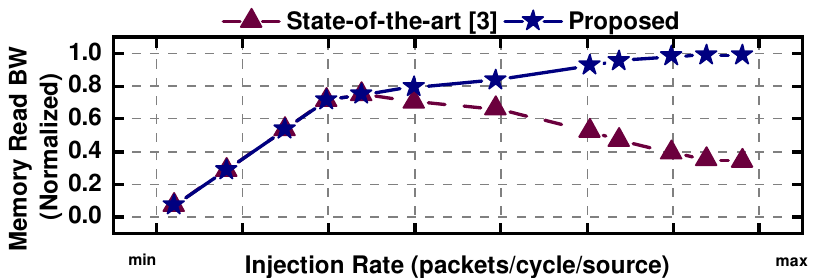}
	\vspace{-3mm}
	\caption{Comparison of memory read bandwidth for 20\% LLC hit rate. Higher memory read bandwidth indicates less NoC congestion.}
	\vspace{-3mm}
	\label{fig:memBW}
\end{figure}

\begin{figure}[t]
	\centering
	\vspace{-2mm}
	\includegraphics[width=0.9\columnwidth]{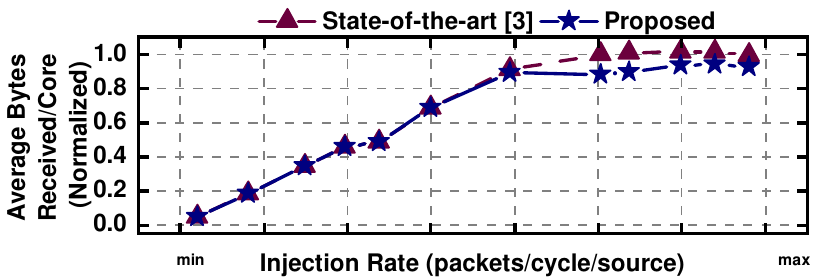}
    \vspace{-4mm}
	\caption{Comparison of average bytes received per core for 70\% LLC hit rate. }
	\vspace{0mm}
	\label{fig:avgBytes}
\end{figure}

\vspace{-3mm}
\subsection{Comparison of Memory Read Bandwidth}
\vspace{-1mm}

This section compares the memory read bandwidth achieved by the state-of-the-art approach and our technique.
Memory read bandwidth measures the average number of requests fetched from the memory controller in case of a cache miss.
The percentage of missed packets with our proposed congestion control technique also significantly increases the memory read bandwidth.
Figure~\ref{fig:miss_perc} shows the 70\% LLC hit rate comparison between state-of-the-art and proposed techniques.
Both techniques result in equal memory read bandwidth at lower injection rates.
However, with increasing injection rate, the memory read bandwidth decreases significantly with a state-of-the-art congestion control technique.
Our proposed congestion control technique keeps the memory read bandwidth at a certain level, even at a higher injection rate.
The highest improvement seen in memory read bandwidth is 190\%.
On average, the proposed technique achieves a 64\% improvement in memory read bandwidth compared to state-of-the-art methods.

Transactions with an LLC miss take significantly longer than those with an LLC hit due to off-chip memory access.
Therefore, the requests with LLC miss stay longer in the queue, reducing the total volume of data received per core (Bytes/core).
However, our technique reduces congestion in the NoC and hence total volume of data received per core does not decrease drastically despite 
 substantial increase in memory read bandwidth. 
Figure~\ref{fig:avgBytes} shows that our proposed approach results in slight (4\% on average) reduction in average bytes received per core although it completes 3.1$\times$ more transactions with LLC miss.

\begin{figure}[t]
	\centering
	\vspace{-8mm}
	\includegraphics[width=0.9\columnwidth]{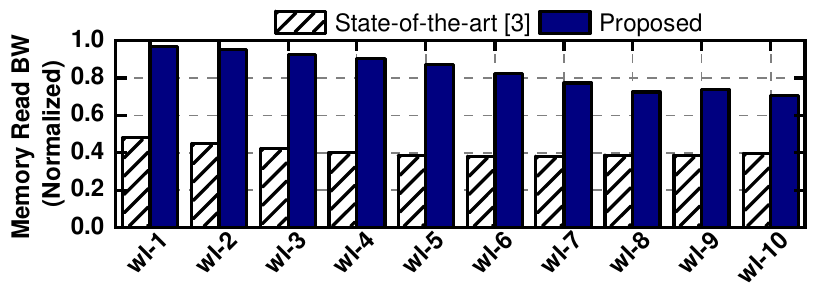}
	\vspace{-6mm}
	\caption{Memory read bandwidth comparison with realistic workloads.} 
	\vspace{0mm}
	\label{fig:memBW_mixtrace}
\end{figure}

\textbf{Results with varying injection rates:} So far, we have shown the results when the cores inject with a fixed injection rate for the entire duration.
However, real applications may have different phases, and in each phase, cores may inject at different injection rates.
Therefore, we also perform experiments with synthetic workloads having different injection rates.
Specifically, in each workload, we consider four different injection rates.
Figure~\ref{fig:memBW_mixtrace} shows the comparison of memory read bandwidth for ten such workloads executing with
70\% hit rate.
Our proposed congestion control technique achieves up to 
114\% improvement in memory read bandwidth compared to the state-of-the-art method.
On average, the proposed congestion control technique shows a
106\% improvement in memory read bandwidth for these realistic workloads.

\vspace{-1mm}
\subsection{Hardware Overhead Analysis} \label{sec:hw_ov}
\vspace{-1mm}

We implemented the RTL for the local condition at all sources and the feature computation as well as the decision tree at all ingress of the NoC.
Then, we synthesized the RTL using Synopsys Design Compiler with 45 nm technology from TSMC.
To have a fair comparison, we scaled the area and power values to 14 nm technology (using the technique described in~\cite{sarangi2021deepscaletool}) since the state-of-the-art SkyLake SoC is fabricated with 14nm~\cite{tam2018skylake}.
We observe that our proposed congestion control technique consumes only 0.01 mm$^2$ of area and 2.2 mW of power.
The total area of SkyLake SoC is 694 mm$^2$ and it consumes power in the order of 10W~\cite{tam2018skylake, WikiChip}.
Hence, our proposed technique incurs negligible overhead (less than 0.01\%) both in area and power.
Therefore, the technique results in significant reduction of NoC congestion with negligible hardware overhead.

\vspace{-1mm}
\section{Conclusion and Future Work} \label{sec:concl}
\vspace{-1mm}

State-of-the-art NoC congestion control techniques are reactive, i.e., can detect NoC congestion only after it occurs.
This paper proposes a supervised learning framework along with a time reversal technique to construct a lightweight decision tree.
This decision tree proactively determines whether any given sink node will likely experience congestion or not (before the queue is blocked).
Experimental evaluation shows that the proposed congestion control technique achieves up to 114\% improvement in memory read bandwidth for realistic workloads while incurring less than 0.01\% of overhead.
\vspace{-2mm}



\footnotesize
\bibliographystyle{unsrt}
\bibliography{embedded_refs}

\end{document}